# Procedural Generation of Communication Networks in Power Systems

Xavier Weiss [1,*], Lars Nordström [1], Patrik Hilber [2], and Emre Süren [3]

[1] Division of Electric Power and Energy Systems, KTH, Sweden
[2] Division of Electromagnetic Engineering and Fusion Science, KTH, Sweden
[3] Division of Network and Systems Engineering, Sweden

## Abstract

Power system communication networks enable operators to remotely monitor and control field equipment. The sophistication of these networks is also increasing as operators continue the trend towards digitization, which is beneficial in integrating distributed energy resources. However, as the attack surface increases in size so too does the risk of cyberattacks. The topology, configuration and composition of communication networks is therefore confidential since this can provide information to attackers. As a result, the number of benchmarks available for research purposes is limited.

A tool for procedurally generating communication network topologies is therefore proposed. While primarily intended as an enabler for public research into communication networks, this tool also allows general insights to be gained into the effect of communication network design on the vulnerability of networks to cyberattacks. The tool includes the ability to encapsulate network characteristics in JSON specification files, which is demonstrated with example Advanced Metering Infrastructure (AMI), Supervisory Control and Data Acquisition (SCADA) and Wide Area Monitoring (WAM) specification files. The SCADA network generation is then compared to a real-world case. Finally, the effect of network redundancy on the networks' cyber resilience is investigated.

## 1 Introduction

### 1.1 Motivation

Communication networks enable grid operators to infer and influence the state of the physical electrical grid. These networks are typically designed to transfer sensor data from devices monitoring specific equipment up to the control center. In response to the sensed information, they can also be utilized to send commands down to reconfigure or control that equipment. The exact characteristics of a network can vary widely depending on what type of equipment it is designed for, what communication components it is composed of, and what technical, physical, or social requirements it must satisfy. Procedural communication network generation is a method for generating a large population of network topologies. By examining this population, promising candidate designs can be discovered, novel methods can be validated, and confidential real-world examples can be kept secret.

The number of possible communication network topologies in power systems far exceeds the actual number of communication networks. Methods designed or trained to work on this limited set of real-world examples may therefore fail when new communication network designs emerge. By providing more example networks, procedural generation can therefore aid in validating the generalizability of state-of-the-art methods for analyzing, controlling or compromising power system communication networks. Procedural generation also benefits data-driven methods, such as machine learning, which tend to perform better when given more diverse and a greater amount of training data.

Smart grids may be characterised by a greater degree of decentralization and digitization than a traditional grid. The trend in this direction is motivated by a need to accommodate non-dispatchable renewable generation through flexible operation rather than energy storage. However, smart grids require more extensive communication networks which can monitor or control a greater range of equipment. As a result, they are more vulnerable to cyberattacks. To estimate this, however, is challenging since the susceptibility of these networks to cyberattacks will depend on specific network configurations, security protocols and which vendors the operator purchased their components from. Since this information is typically confidential, procedural generation is particularly applicable to power system cybersecurity studies.

### 1.2 Previous Work

As illustrated in Fig. 1, numerous protocols and standards are available that can aid in designing power system communication networks. Distributed Network Protocol (DNP3) is, for instance, used in SCADA systems to provide error checking, time synchronization and response/request messages. IEC 61850 [2] provides a framework for substation automation and communication in terms of logical nodes and mappings to specific communication protocols such as the Sampled Values protocol or the Manufacturing Message Specification (MMS) protocol. However, these standards may be updated or replaced and provide greater granularity than is necessary to characterize a communication network. For this reason, standards in this work are referred to as a general guide for defining network types, but specific communication protocols are ignored. Topology generation is an older, well established, technique in internet research. Initially, internet-like topologies were generated by purely random assignment of links between a fixed number of nodes [3], which can link any node to any other node regardless of the distance between them. The Waxman method [4] attempts to capture the prefer-ence to connect nearby nodes to one another by weighting link generation based on the Euclidean distance. Since the internet expanded organically, new nodes tended to connect to existing, well-connected nodes, which motivated the application of power laws to capture the heavy tail in node degree / connectivity [5]. However, as observed by [6] this is only true at the edge of the network since the network core tends towards high connectivity because of the economics / performance demands. BRITE [7] combined some of these techniques in an extensible toolbox and identified some requirements for topology generation such as efficiency and extensibility. For the power system domain efficiency is also needed to scale from local distribution system networks to regional and even country-spanning networks. Extensibility, such as accounting for new types of devices and network types is also needed as telecommunication advances.

*Correspondence: xavierw@kth.se

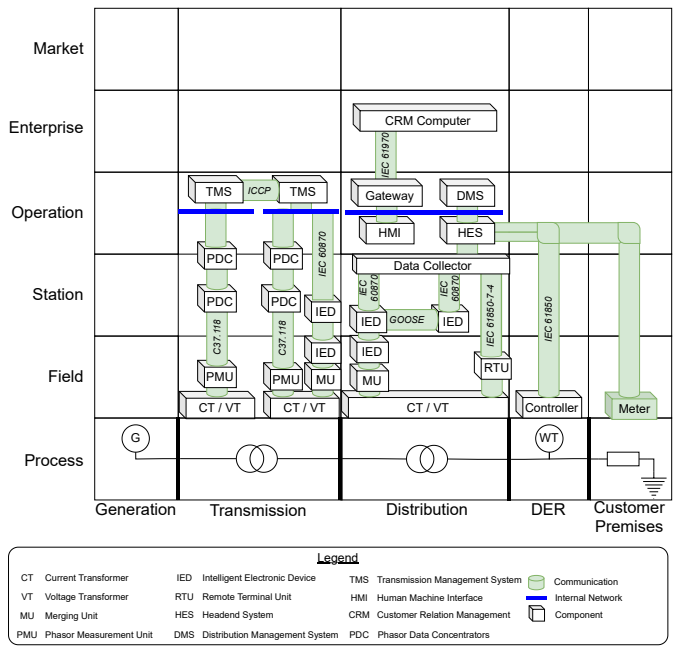

Figure 1: Smart Grid Architecture Model [1] with various communication protocols

To capture the hierarchical structure of the internet recursive strategies are needed. The tiers method does this by dividing the internet into Local, Metropolitan and Wide Area Networks (LAN, MAN, WAN) and connecting them via a minimum spanning tree algorithm. Others have clustered nodes into sub-networks [8] or drawn links between stub-domains, transit domains and hub domains [3]. IGen extends this by considering geographical constraints. In the case of the internet, public topologies are known [9] and their statistics can be used as an input for generating and validating network topologies [10]. However, in the field of power systems such topologies are not generally public which makes validation through direct comparisons - e.g. to case studies - and statistical comparisons - e.g. to node degree - challenging. Instead, a more transparent heuristic approach is justified so that experts can themselves judge whether the generated topologies are useful for their application. Lessons can nevertheless be drawn from later internet topology generation tools to capture the hierarchies present in power system communication networks, such as clustering devices together.

Procedural network generation has been applied to other forms of infrastructure, such as road networks. [11] interactively generate street graphs based on user-defined network patterns which can be edited using brushes. [12] avoid user interaction by automatically finding intersections, crossings, roundabouts and city blocks based on a center-line estimate of where real-world roads are. [13] generates a 3D road network for virtual driving consisting of highways, secondary roads and rural roads based on terrain data and specifications about the environment (rural, city, etc.). These works demonstrate the utility of procedural techniques in modelling infrastructure, but either rely on real-world baselines or human interaction to produce their graphs. In reality, real-world data may be too sensitive to rely on. For instance, communication network designs may reveal potential vulnerabilities to cyber attackers. Additionally, generating networks through human interaction is slow and scales poorly for data-heavy applications, such as machine learning. Instead such expert knowledge can be used to automate and constrain the types of network graphs that can be generated.

Communication network design in power systems has generally been treated as an optimization rather than a network generation problem. [14] assigns communication channels to sensors and controllers by minimizing interference using the Hungarian algorithm. [15] instead minimize delay and packet loss while maximizing reliability using a genetic algorithm. [16] optimize data transmission between controllers to help dampen small-signal instabilities. However, real-world power system communication networks are not necessarily built based on numerical optimization, since social, physical and technical constraints may not be quantifiable. Hence instead of choosing a network to optimize an objective, various networks can be generated purely based on what grid equipment requires a connection.

Network generation has been applied to the physical equipment in the electrical grid. [17] generate random electrical grid topologies to enable statistical analysis of future communication network designs. [18] use a minimal spanning tree algorithm to synthesize a transmission network that connects load clusters with generation. Finally, [19] use a loop-oriented tree-growth algorithm to generate foam-like electrical grids with nodes and powerlines. Such synthetic grids can augment existing benchmarks and real-world datasets to validate the generalizability of algorithms or models. However, as pointed out in [20], these synthetic grids may differ in graph properties from their real-world counterparts. For this reason, domain knowledge is required when synthesizing networks to ensure the resulting topologies are realistic.

### 1.3 Contributions

Based on the considerations above, this work therefore introduces a procedural generation algorithm for synthesizing tree-like communication networks for use in power systems. This algorithm includes the following contributions:

1. Generation of tree-like communication networks for Supervisory Control and Data Acquisition (SCADA), Wide Area Monitoring systems (WAMs) and Advanced Metering Infrastructure (AMI).

2. Extensibility to other communication network types through the use of human-readable JavaScript Object

Notation (JSON) specification files that encapsulate knowledge specific to the network type.

3. Integration of communication networks with power systems by assigning communication devices to specific physical equipment based on specified conditions.

4. Applicability to aggregated (i.e. higher voltage) grids by assigning multiple communication devices to aggregated equipment based on specified splitting criteria.

## 2 PROCESS

Procedural methods, while most commonly seen in the video game and 3D animation industries, have been used to plan infrastructure - such as in the design of roads in [21] or cities in [22]. In this work we apply procedural generation to the creation of communication network topologies using a recursive algorithm with randomized parameters. By generating many variants of a communication network an existing design can be compared to alternatives, or a new design can be chosen based on the susceptibility of the topology to cyberattacks.

The procedural generation algorithm, shown in Algorithm.1, takes 3 external inputs: a physical grid model (**grid**, optional), a JavaScript Object Notation (JSON) specification file (**spec**) and a set of hyperparameters. The specification file is outlined in Figure.2 and provides information on the types of device, aggregator and root nodes in the network. At the time of writing, example specifications are provided for generic, SCADA, smart meter and WAM networks. To be added are protection networks, wireless communication and the ability to simulate communication protocols. The output of Algorithm.1 is an abstract representation of a communication network which models how communication components are connected to share information without making any specific assumptions on what protocols are used to share that information.

The grid is a PandaPower [23] network containing a representation of the physical equipment in an electrical grid, such as transformers or lines. If provided, communication network devices are mapped to components in the PandaPower grid based on the specifications. Features specific to the physical grid representation as well as the details of how to evaluate whether physical equipment meets conditions set in the specifications have been omitted from Algorithm 1 for clarity.

'Random()' chooses a device or aggregator category either uniformly or based on a weighting in the specifications (e.g. 30% from vendor A and 70% from vendor B). If the equipment is selected based on conditions, this weighting is re-normalized after filtering out incompatible equipment. 'get_n_splits()' returns the number of devices for a specific piece of equipment, which depends on the conditions specified in the specifications (e.g. an aggregate load of 10kW might consist of ten 1kW smart meters). nChildren is either randomly distributed about 0 by some integer deviation, or is equal to the total number of components if the communication network topology is specified as flat. If inside the algorithm the number of children is negative, this means those components will be connected 1 level higher in the hierarchy.

The specifications in Figure.2 include the Cumulative Distribution Functions (CDFs) that describe the effort required to try and

**Algorithm 1** Pseudocode of procedural hierarchical network generation algorithm

```
1:  components = [ ]
2:  if Length(components) = 0 then          ▷ Leaf Nodes
3:      for i = 0, ..., N do
4:          category = Random(spec["device"]["categories"])
5:          nSplits = get_n_splits(grid, category, i)
6:          for split = 0, ..., nSplits do
7:              device ← create_device(category)
8:              components.append(device)
9:  else if Length(components) > 1 then     ▷ Internal Nodes
10:     n = 0
11:     remaining = [ ]
12:     while n ≠ Length(components) do
13:         nChildren = get_n_children()
14:         if nChildren ≤ 1 then           ▷ Skip some child nodes
15:             remaining.append(components[n : n + nChildren])
16:         else
17:             category = Random(spec["aggregator"]
18:                                     ["categories"])
19:             aggregator ← create_aggregator(category)
20:             for i = n, ..., n + nChildren do   ▷ Edges
21:                 connect(aggregator, components[i])
22:                                                ▷ Siblings
23:                 if i ≥ 1 & sib2sib = "adjacent" then
24:                     connect(components[i − 1],
25:                             components[i])
26:                 else if i ≥ 1 & sib2sib = "all" then
27:                     for j = n, ..., n + nChildren do
28:                         connect(components[j],
29:                                 components[i])
30:             remaining.append(aggregator)
31:         n = n + nChildren
32:     components = remaining
33: else if Length(components) = 1 then     ▷ Root Node
34:     root ← create_root(spec["root"])
35:     connect(root, components[0])
36:     STOP                                 ▷ End Recursion
37: else
38:     GOTO LINE 2                          ▷ Recursion
```

compromise the communication network component and the probability that the attempt succeeds. These CDFs are inspired by the Meta Attack Language (MAL) [24], but unlike MAL are kept generic to not over-specify the system, since actual effort and probability distributions are often unknown or poorly estimated. However, it is possible to state that the root node - typically the control center - is usually the most hardened, aggregators are less hardened and leaf nodes - especially in the case of smart meters - are least hardened. This abstraction allows conclusions to be made about the relative strength of differential topologies irrespective of specific events in cyberspace, such as the release of a new zero day exploit. Nevertheless, the defence CDFs can be changed in their minutiae by a DSO to reflect their expert knowledge if more specificity is needed.

Algorithm.1 takes several hyperparameters. As illustrated in Figure.3, these can have a strong impact on the network topology. The hyperparameters include:

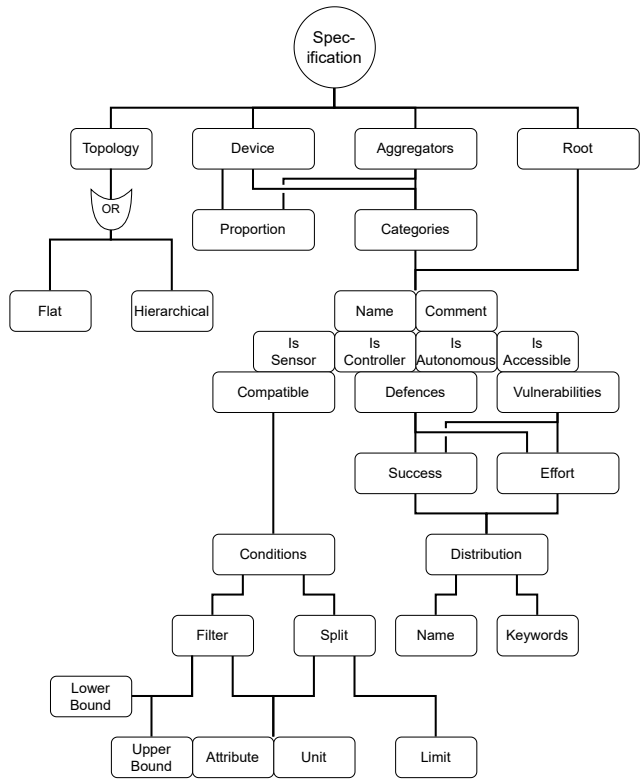

Figure 2: Breakdown of the components of the JSON specification files.

**Sibling To Sibling Communication:** If enabled, lateral communication connections exists between children of the same parent node. This can for instance represent Intelligent Electronic Device (IED) to IED communication protocols like GOOSE, though these are not yet explicitly modelled. By default a node can reach all other siblings with the same parent.

**Flat:** If True, treat the communication network hierarchy as flat, where all devices connect to a single aggregator.

**Children Per Parent:** Average number of child nodes per parent node.

**Children No. Deviation:** Random deviation in the number of children per parent.

**No. of Entrypoints:** Number of randomly allocated entrypoints in the network where attackers can attempt to enter the network. A value of 1 is used in this paper.

**Compatibility:** Provided by specifications, indicates which types of PandaPower components the communication device can be placed at. Additional conditions may apply, such as a minimum voltage level.

**Proportion:** What proportion of the population of devices belong to each category (e.g. Controller vs Sensor in Fig.3). If no specification is provided, this determines the overall proportion, otherwise it is re-normalized based on which devices meet the compatibility conditions.

**Seed:** A random number generation seed, used for reproducibility.

A distinction can be made between the parameters of the algorithm, such as the number of children of a specific aggregator, and the hyperparameters of the algorithm, such as the average number of children per aggregator. The hyperparameters influence the space of parameters, however, if the algorithm is run multiple times with different seeds it is unlikely to generate the same network twice unless the hyperparameters are overly specific.

Algorithm.1 is able to generate a wide variety of communication networks. By attacking each variation in network topology with a large number of randomized attacks, a profile can be built for each network variant. To improve the relevance of this approach, the following subsections describe the specifications for common distribution system communication networks.

## 3 Communication Networks

As illustrated in Fig.4 there are broadly 3 types of communication networks in power systems: AMI, WAM and SCADA. In this section procedurally generated example networks for each are presented and discussed from a cybersecurity perspective.

### 3.1 Advanced Metering Infrastructure

Advanced metering infrastructure enables utilities to bill customers automatically and gain insight into the behaviour of customers at a distribution level. Through two-way communication they can also be employed for demand response, such as through variable tariffs, or for load shedding by disconnecting individual customers. According to [25], the data from smart meters is communicated either through wired connections, such as power line communication, or wirelessly, such as through WiFi. As shown in Fig.5, this data ultimately connects to the DSO's control room, but is usually separate from traditional SCADA systems.

Communication network components can be provided by many different vendors, each with its own security policies. This reality is demonstrated in Fig.5 with 2 smart meter vendors. These are often installed at scale by specific utility companies, for instance when a new neighborhood is built. Consequently, attacks on smart meters are often scalable - since a single exploit can affect all smart meters of that type. As described in [27], attackers have already used smart meters to construct botnets for Distributed Denial of Service (DDoS) attacks. To impact power system operation smart meters could be used for False Data Injection (FDI) attacks or - in the case where remote demand response or load shedding is possible through the smart meter - coordinated changes in demand can hypothetically be used to destabilize the grid.

### 3.2 Supervisory Control and Data Acquisition (SCADA)

SCADA systems are two-way communication networks dating back to the 1960s [28] that allow DSOs to both observe what happens in their grid, and perform control actions in response. According to [29], SCADA for power systems typically consist of field devices, such as Remote Terminal Units (RTUs), that connect to a SCADA server via a Communication Front End (CFE). The SCADA topology is typically flat, though there may be some hierarchical structure to substation-level automation.

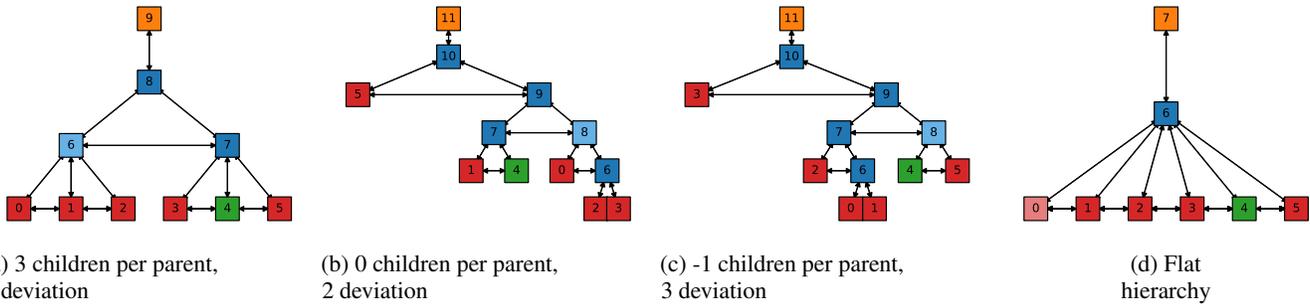

(a) 3 children per parent, 0 deviation

(b) 0 children per parent, 2 deviation

(c) -1 children per parent, 3 deviation

(d) Flat hierarchy

Figure 3: Generic communication network topologies with 6 devices using the seed 7977 and default specification file. All communication networks are modelled in terms of devices (leaf nodes), aggregators (internal nodes) and a control center (root node).

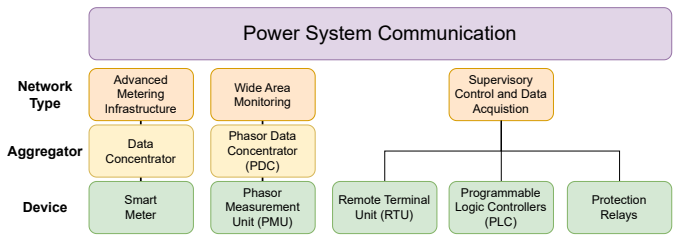

Figure 4: Taxonomy of Communication Networks

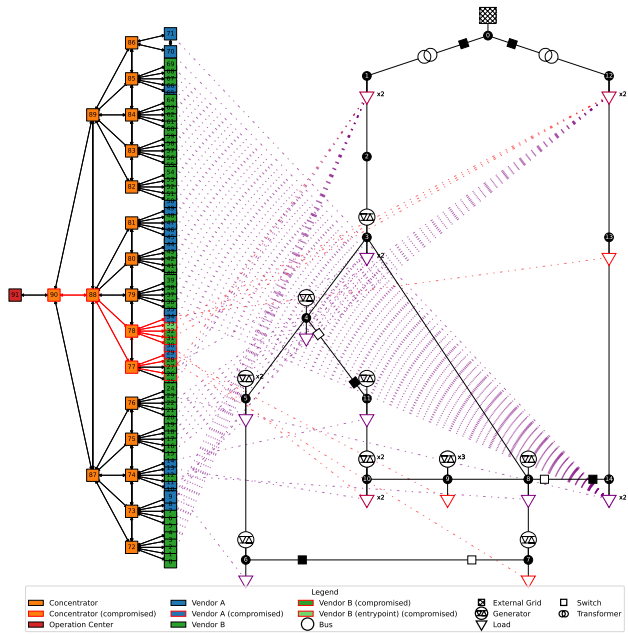

Figure 5: Example AMI communication network with 2 vendors for CIGRE MV grid [26]. Purple equipment have communication device(s) installed and Red equipment has compromised device(s) attached.

An analysis of 20 years of SCADA attacks in [30] points out that remote maintenance interfaces, typically carried out over the internet, are particularly vulnerable to attacks.

As seen in Fig.6, the developed SCADA model uses a single entrypoint to represent a remote maintenance connection. To avoid unnecessarily modelling all possible varieties of devices and situations that SCADA covers, a simplified model with Fault Passage Indicator (FPIs) and RTUs as field devices is considered. Since the network is flat, the procedural generation algorithm will always produce the same topology. However, due to the flat hierarchy the SCADA system shown has a single point of failure - the CFE - which if compromised could allow attackers to intercept, manipulate, or execute any commands.

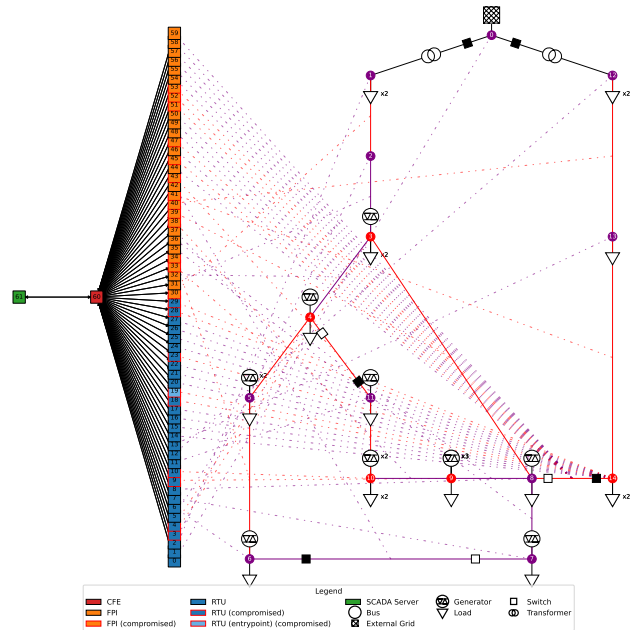

Figure 6: Example SCADA communication network for CIGRE MV grid [26]

### 3.3 Wide Area Monitoring (WAM)

WAMS is designed to address the time-skew and low-update frequency seen in SCADA. As described in [31], it is designed to communicate high frequency, time-synchronized phasor measurements. Phasor Measurement Units (PMUs) measure time-synchronized voltage/current magnitudes and angles at locations in the grid. Layers of Phasor Data Concentrators (PDCs) then aggregate measurements from these PMUs, or other PDCs, until they reach a central PDC. Since WAMs enables real-time processing, it can be used to detect angle differences, low frequency oscillations, islanding, voltage stability issues, and more.

The structure of a WAMS network shown in Fig.7 is similar to the AMI network in Fig.5. However, unlike smart meters, PMUs and PDCs are typically not found at a distribution level due to their high cost. Since power flows can be deduced from WAMS measurements, compromised assets can reduce observability and potentially interfere with control schemes.

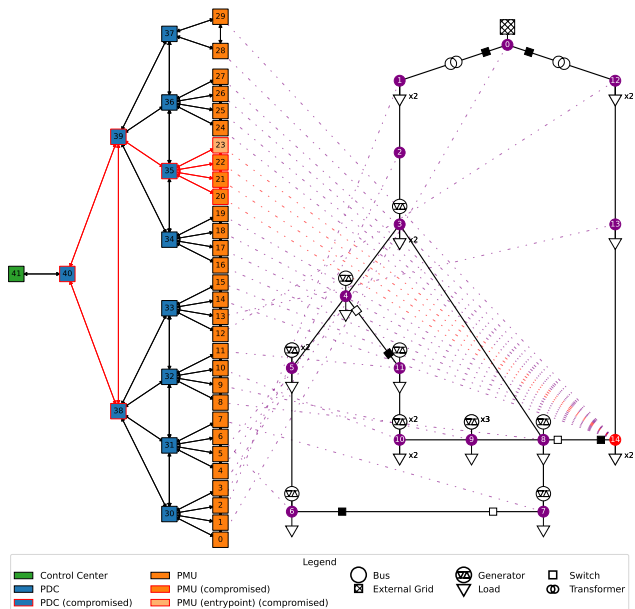

Figure 7: Example WAMS communication network for CIGRE MV grid [26]

## 4 Validation

### 4.1 Real-World Comparison

The procedural networks outlined in Section.3 were designed based on a theoretical understanding of how communication networks in power systems are typically arranged. However, the practical arrangement of real-world networks may differ. To see how these might differ, a comparison is therefore made here to the communication infrastructure for a LV/MV system in [32].

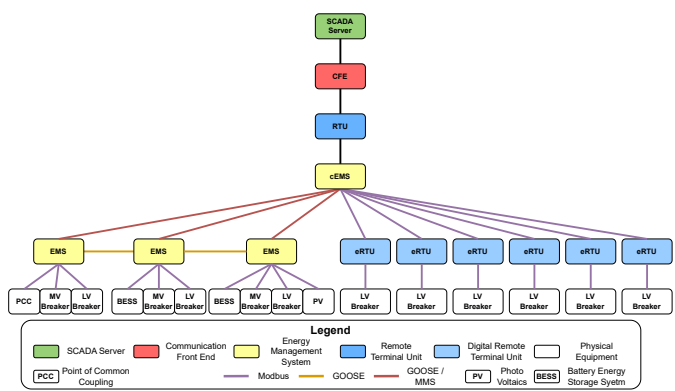

Figure 8: Diagram of a real-world LV/MV Communication Network [32]

Fig.8 shows that the LV communication network has a tree-like structure similar to those Section.3. Like the SCADA example, it has a flat hierarchy and includes sibling-to-sibling communication between some of the Energy Management Systems (EMSs). However, Algorithm.1 neglects both the protocols and mediums for communication between devices in the network. Likewise, because it relies on uniform random numbers, lopsided configurations are unlikely to occur. The population of possible networks generated by Algorithm.1 may therefore not reflect the population of real-world networks, since these are influenced in design by practical constraints rather than random effects. Thorough validation of this will, however, require access to a larger amount of real-world examples. A possible alternative is to gather expert feedback on the generated topologies, such as via a survey, in order to get qualitative data on the accuracy of the topologies. However, due to a lack of ready access to experts, this is left to future work.

### 4.2 Cybersecurity

One application for procedurally generated communication networks is cybersecurity. Here the effect of network generation hyperparameters on the vulnerability to cyberattacks are checked to see whether changes to network structure have a sensible impact. Since access to real cybersecurity data is not feasible, further verification is currently not considered.

A brief summary of how the cyber security is modelled is provided here, a full explanation is outside of the scope of this work however for more details please see []. With the assumption that the devices (leaf nodes) are easiest to compromise, aggregators (internal nodes) are harder to compromise and the control center (root node) is the hardest to compromise, each element in the network is assigned both a compromise and effort probability distribution. A Monte Carlo approach is then taken where an attacker is assigned a random entrypoint, samples the probability distributions to see if it successfully compromises it and how long it takes, and then chooses its next target randomly from the adjacent nodes to the node it just compromised. This is done recursively until either the entire network is compromised, the attacker runs out of budget, or reaches a dead end. Since the starting point of the attack can influence the possible compromise distribution, the entrypoint is varied such that each of the 116 devices is a starting point for 1000 of the attacks. This does not capture more sophisticated strategies of real attackers, who most likely do not randomly compromise nodes, nor the presence of zero-day vulnerabilities. However, assuming most attacks are opportunistic rather than targeted, by running enough simulations the network's general susceptibility to cyberattacks can be identified.

#### 4.2.1 Children per aggregator

Redundancy is modelled as the average number of children per aggregator. The most redundant network is one where each aggregator has 2 children. The least redundant network has all devices connected directly to one aggregator. Since greater redundancy requires more aggregators, and thus more components in the communication network, it is more expensive for a DSO to have a highly redundant setup.

Since the role of aggregators is to aggregate data from lower levels, which is generally a unidirectional process, we consider the compromise distribution from the perspective of how many devices, not aggregators, are compromised. That is we assume only compromised devices have a direct operational impact. This reflects the purpose of the aggregators, which typically pass information from one level to the next and do not necessarily have the credentials to issue or imitate commands from the control center nor to prevent commands from reaching the devices. Compromised aggregators can, however, have indirect opera-tional impacts that are not modelled here. For instance, they may reduce pass on inaccurate readings to the control center that leads them to issue incorrect commands. In the simulations here an aggregator is simply a bottleneck for an attacker that must be compromised in order to reach its children.

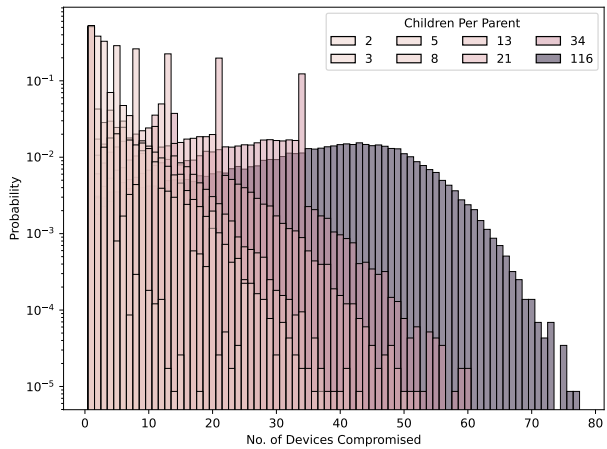

Figure 9: Variation in compromise distribution with differing number of children per aggregator.

Figure.9 shows the least (116 children per aggregator) and most (2 children per aggregator) redundant compromise distributions, with varying levels of redundancy in between. For each case a total of 116,000 attacks are simulated, 1000 for each possible entrypoint. To ensure the networks remain consistent, no variation is allowed in the number of children per aggregator - except for rounding when the device count does not split exactly. Consequently, the only randomness in the Monte Carlo simulation is the attack path taken and the randomly sampled effort required. As is seen, the network tends towards fewer compromises with high redundancy (few children per aggregator) compared to lower redundancy (many children per aggregator). Since we are examining the compromise distribution of devices, having more aggregators not only constraints which parts of the network are reachable by the attacker at any point (thus making a dead-end scenario more likely) but also means more effort must be expended compromising aggregators before the devices can be reached. Hence the results follow intuition on redundancy, with greater redundancy yielding a more cyber-resilient network. This confirmation provides some credence that the procedural generation is useful as a research benchmark.

#### 4.2.2 Sibling-to-sibling communication

Children of the same parent in the communication network can either only have connections with their parent (sibling-to-sibling: False), only with directly neighbouring siblings (sibling-to-sibling: Adjacent), or all siblings (sibling-to-sibling: All). This provides more possible attack paths for attackers.

Figure.10 shows that increased sibling-to-sibling communication leads to a rightwards shift in the compromise distribution. However, since the effort could be interpreted as exhausting the patience or resources the attacker is prepared to expend to compromise a component, if the attackers have previously tried and failed to compromise the parent, they will not attempt to attack it again from the sibling node. For this reason, since the parent can act as a bottleneck for reaching other parts of the network, the effect of lateral movement is limited. Note that lateral movement does not exist between nodes of the same level, but with different parents. This reflects that devices may be connected with sibling-to-sibling communication in groups, such as within a substation. Hence the effect of sibling-to-sibling communication, like redundancy, follows intuition - more connections means more possibilities to compromise and hence a more vulnerable network.

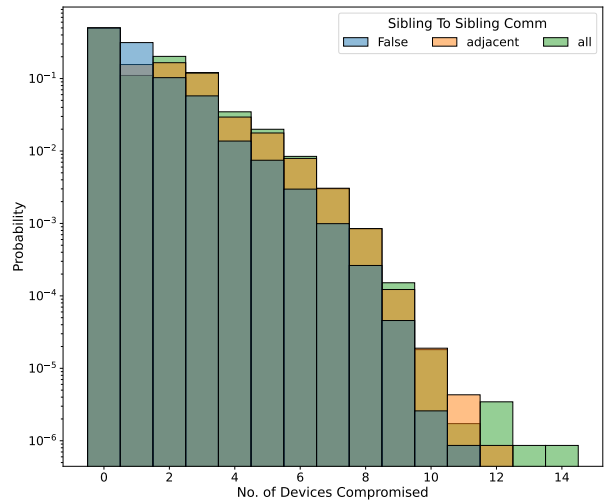

Figure 10: Variation in compromise distribution with different types of sibling-to-sibling communication.

### 4.3 Scalability

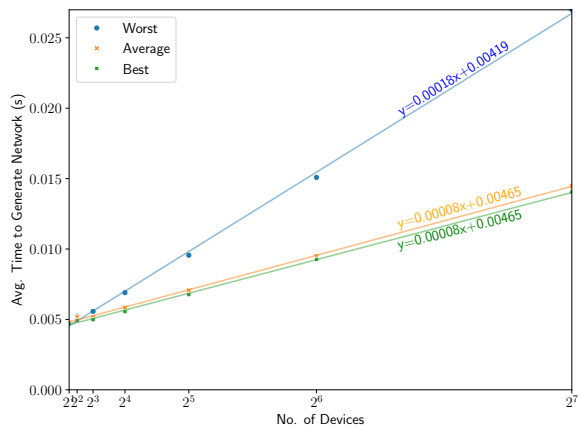

Figure 11: Average computational time taken to generate one procedural network, based on 10,000 single-thread trials.

Since the generation of a network topology does not depend on the other networks that have been generated, Algorithm.1 is embarrassingly parallelizable. To keep the scaling comparison simple, however, only a single core on an i7-77000K CPU is used to generate 10,000 networks with worst case (most redundant → most components), average case (randomized no. of children per aggregator) and best case (least redundant → least components) network complexities. The result, as shown in Figure.11 is a clear linear increase in computation time with network size. Based on the lines shown, fit using linear regression, it would take approximately 80 seconds to generate a network containing 1 million devices (a quick test of this took ∼ 77.1 seconds). Given that such networks can be generated offline and typically only need to be generated once, the authors' consider this to be sufficiently efficient (see [7]) for topology generation - especially if the additional speed-up from parallelization is included.

## 5 Conclusion

The proposed procedural generation algorithm is able to generate a variety of tree-like power system communication networks. These networks are designed to be abstract representations of the layout of a typical communication network, since they are used to evaluate cybersecurity. Domain experts can specify new network types through a specification file that specifies the cybersecurity characteristics of components in the network, the proportion of different device types, and the rules for determining where and how many devices are placed. Example network types are provided for Supervisory Control and Data Acquisition (SCADA), Advanced Metering Infrastructure (AMI), Wide Area Monitoring (WAM) and generic networks. The generic network type was then checked in the cybersecurity domain by varying the hyperparameters of the algorithm. The results match expectations, with increased redundancy yielding networks with fewer compromises. Future work may validate this algorithm through expert feedback and potentially extend it to satisfy reliability, cost, or geographical constraints. In addition, by including communication network protocols it would become possible to simulate network activity, information loss, and latency for research testbeds.

## 6 Acknowledgments

Funded by Energimyndigheten, project number: P2022-00692